\documentclass[12pt]{article}
\usepackage{amsfonts}

\title{On the construction of ${\cal N}=4$ SYM effective action beyond leading low-energy
approximation \footnote{Talk given at International Workshop
"Supersymmetries and Quantum Symmetries" - SQS'03 (Dedicated to
the 75th anniversary of the birth of V.I.Ogievetsky); JINR, Dubna;
July 24-29, 2003 }}

\author{A.T. Banin\footnote{atb@math.nsc.ru},
N.G. Pletnev\footnote{pletnev@math.nsc.ru}}

\date{{\it
Institute of Mathematics, Novosibirsk, \\ 630090, Russia}}%\\
\begin{document}

\maketitle

\begin{abstract}
A problem of the hidden ${\cal N}=2$ supersymmetry deformation for
next-to-leading terms in the effective action for ${\cal N}=4$ SYM
theory is discussed. Using formulation of the theory in ${\cal
N}=2$ harmonic superspace and exploring the on-shell hidden ${\cal
N}=2$ supersymmetry of ${\cal N}=4$ SYM theory, we construct the
appropriate hypermultiplet-depending contributions for $F^6$ term
in the Schwinger-De Witt expansion of the effective action. The
procedure involves deformed hidden ${\cal N}=2$ supersymmetry and
allows one to obtain self-consistently the correct ${\cal N}=4$
supersymmetric functional containing $F^{6}$ among the component
fields.

\end{abstract}
\thispagestyle{empty}

\newcommand{\be}{\begin{equation}}
\newcommand{\ee}{\end{equation}}
\newcommand{\bea}{\begin{eqnarray}}
\newcommand{\eea}{\end{eqnarray}}

\section{Introduction}
Different aspects of low-energy string dynamics and of the AdS/CFT
 correspondence \cite{1} can be studied in terms of the
quantum field theory effective action. Due to the proposition put
forth in (see discussed in greater detail \cite{2, 3, 4}), the
superconformal version of the Born-Infeld (BI) action is expected
to be considered as the result of summing up leading and
subleading terms in the quantum effective action of the ${\cal
N}=4$ SYM theory in the Coulomb branch. Some terms of the
effective action expansion should be constrained by new
non-renormalization theorems.

In the off-shell superfield formulation of the SYM and
supergravity theories, the supersymmetry must be realized linearly
on physical fields and on an infinite set of auxiliary fields so
that the supersymmetry transformations are independent of the form
of the action. But in the on-shell formalism, the supersymmetry
transformation is realized non-linearly. When obtaining higher
derivative contributions to the effective action preserving
extended supersymmetry, we must obtain self-consistently the
deformations of the classical supersymmetry transformation rules
order by order and, simultaneously, construct the
supersymmetry-invariant higher-order terms in the action:
$(\delta_0+\sum_n\delta_n)(S_0+\sum_n S_n)=0$. Here $\delta_{0}$
is classical supersymmetry transformation, $S_{0}$ is classical
action and $\delta_{n}, S_{n}$ are the quantum deformations of
transformation law and higher derivative corrections to the
classical action respectively.  It is hardly possible to compute
in a closed form the full derivative dependence of the effective
action (see for review \cite{5}). Therefore, what we can do in
this situation is, relying on particular results known, to list
all the supersymmetry invariants as well as deformed
transformation rules with a given number of derivatives. For the
well-known leading potential $\propto F^4$ in the vector field
strength sector (see for references and a recent progress
\cite{5}), the problem of constructing the full ${\cal N}=4$
superinvariant has been solved in \cite{6}.

There exist many different approaches that are used for the
construction of the supersymmetric higher-derivative string
effective action \cite{7, 8, 9}. In the instructive paper
\cite{10}, an off-shell ${\cal N}=3$ supersymmetric extension of
the Abelian $D=4$ BI action was constructed starting from the
action of supersymmetric Maxwell theory in ${\cal N}=3$ harmonic
superspace \cite{11}. The ${\cal N}=3$ superfield strength
contains combination of the auxiliary field and the gauge field
strength. The nonlinearity in the ordinary gauge field strength
arises in its full form as the result of elimination of these
auxiliary fields with the aid of their nonlinear equations of
motion. This is different from ${\cal N}=1, 2$, where
superextensions of the bosonic BI action are fulfilled in each
order of expansion in powers of the Maxwell field strength.

The purpose of the present paper is to consider a possible
self-consistent way to find hypermultiplet dependent complements
and the correspondent deformed hidden supersymmetry
transformations, which are needed for manifestly ${\cal N}=4$
supersymmetric next-to-leading terms in the ${\cal N}=4$ SYM
theory effective action. The while such an approach is useful for
the resolution of existence problem concerning higher-derivative
invariants, obtaining fully explicit expressions is extremely
cumbersome due to the enormous number of terms and the problem of
dealing witch partial integration.

\section{On hidden ${\cal N}=2$ invariance of the $F^6$ term in ${\cal N}=4$
SYM theory}

In order to construct subleading ${\cal N}=2$ hidden invariant
terms in the derivative expansion of the effective action, which
depend on all the ${\cal N}=4$ SYM multiplet fields, one can apply
the Noether procedure with classical supersymmetry transformation
modulo to boundary terms and free equations of motions
\cite{12},\cite{6}
\begin{eqnarray}
&\delta_0 {\cal W}={1\over
2}\bar{\varepsilon}^{\dot{\alpha}\,a}\bar{D}^{-}_{\dot{\alpha}}q^{+}_{a},
\quad \delta_0 \bar{\cal W}={1\over
2}{\varepsilon}^{\alpha\,a}D^{-}_{\alpha}q^{+}_{a}\label{1},&
\\ &\delta_0 q^{\pm}_{a}={1\over
4}(\varepsilon^{\alpha}_{a}D^{\pm}_{\alpha}{\cal W} +
\bar{\varepsilon}^{\dot\alpha}_{a}\bar{D}^{\pm}_{\dot\alpha}
\bar{\cal W})~.& \nonumber
\end{eqnarray}
Let us try to find a possible hypermultiplet completion for the
following two-loop term $\propto F^6$ found in \cite{3} with the
supergraph technique in the harmonic superspace:
\begin{equation}\label{2}
\Gamma_{(6|0)} = c_2\int d^{12}z [\frac{1}{\bar{\cal W}^2}\ln
{\cal W} D^4 \ln {\cal W } + \frac{1}{{\cal W}^2} \ln\bar{\cal W}
\bar{D}^4 \ln \bar{\cal W} ]~ = c_2 \int d^{12}z {\cal
L}_{(6|0)}+c.c.,
\end{equation}
where $c_2 = N^2 g^2_{YM}\frac{1}{48\cdot(4\pi)^4}$. This term
consists of two different parts, marked by the fourth powers of
distinct derivatives: $D^4$ or $\bar{D}^4$. This parts should be
studied separately, because the variation rules (\ref{1}) do not
mix them. The variation of first part (\ref{2}) induced by
(\ref{1}) can be written in the form
\begin{equation}\label{3}
\delta_0 {\cal L}_{(6|0)} = -\frac{2q^{+a}}{\bar{\cal W}^3{\cal
W}}(\bar\varepsilon^{\dot{\alpha}}_a
\bar{D}^-_{\dot{\alpha}}\bar{\cal W} + \varepsilon^{\alpha}_a
D^-_{\alpha}{\cal W}) D^4 \ln{\cal W} +
\frac{q^{+a}\varepsilon^{\alpha}_a D^-_{\alpha}{\cal W}}{{\cal
W}\bar{\cal W}^3}D^4 \ln{\cal W}.
\end{equation}
The heart of all problems is the fact that the classical variation
(\ref{3}) of ${\cal L }_{(6|0)}$ generates terms non-symmetric
under the replacement $\varepsilon
\leftrightarrow\bar\varepsilon$.

Further we consider classical transformations $\delta_{0}$ defined
by (\ref{1}) along with their deformations $\delta_{(n|k)}$. The
full deformed transformations are considered as an expansion in
powers of $D,\bar{D}$ as well as in powers of
$X=\frac{-2q^{+a}q^-_a}{{\cal W}\bar{{\cal W}}}$, i.e. $\delta =
\delta_{0} + \delta_{1}(D^4) + \delta_{2}(D^8) + \ldots$ The
subscript $k$ in deformations indicates the power of $X$, e.g.
$\delta_1=\sum \delta_{(1|k)}$. Let us introduce the first
complement to (\ref{2}) in the form
\begin{equation}\label{4}
{\cal L}_{(6|1)} = d_1[X\frac{1}{\bar{\cal W}^2}D^4 \ln{\cal W} +
X \frac{1}{{\cal W}^2}\bar{D}^4\ln \bar{\cal W}].
\end{equation}
Its variation $\delta_{0}^{(q)}{\cal L}_{(6|1)}$ in $q^{\pm}$
cancel the first term in (\ref{3}) if $d_1 = -2$, but the other
part of (\ref{3}) is not cancelled. Since the structure either of
the two functionals (\ref{2}) is not symmetric in respect to
${\cal W} \leftrightarrow \bar{\cal W}$, while the
$\delta^{(q)}$-variation is symmetric, the discrepancy between
variations like (\ref{2}) and (\ref{3}) will always appear in each
step of the variational procedure. To resolve this discrepancy, we
consider the one-loop $\Gamma_{(4)}\propto F^4$ term along with
its well-known hypercomplement \cite{6}:
\begin{equation}\label{5}
\Gamma_{(4)} = c_1 \int d^{12}z\,[\ln {\cal W }\ln \bar{\cal W} +
\frac{1}{2}X + \frac{1}{4\cdot3}X^2 + ...]~,
\end{equation}
where $c_1 = N \frac{1}{(4\pi)^2}$. We know that this term will be
renormalizable by neither the higher loop nor the instanton
contributions. Suppose that the classical hidden supersymmetry is
deformed as follows
\begin{equation}\label{7}
\delta_{(1|0)} {\cal
W}=\frac{\bar{A}}{2}\bar\varepsilon^{\dot\alpha
a}\bar{D}^-_{\dot\alpha}q^+_a\frac{1}{{\cal W}^2}\bar{D}^4\ln
\bar{\cal W},\ \delta_{(1|0)} \bar{\cal
W}=\frac{{A}}{2}\varepsilon^{\alpha
a}{D}^-_{\alpha}q^+_a\frac{1}{\bar{\cal W}^2}{D}^4\ln {\cal
W}\end{equation}
\begin{equation}\label{8}\delta_{(1|0)} q^{\pm}_a =
\frac{1}{4}[B \varepsilon^{\alpha}_a D^{\pm}_\alpha {\cal
W}\frac{1}{\bar{\cal W}^2} D^4\ln {\cal W} + \bar{B}
\bar\varepsilon^{\dot\alpha}_a \bar{D}^{\pm}_{\dot\alpha}
\bar{\cal W}\frac{1}{{\cal W}^2} \bar{D}^4\ln \bar{\cal W} ].
\end{equation}
If the following conditions
\begin{equation}\label{eg1}
c_2 + c_1 \frac{(\bar{A} - \bar{B})}{2} =0,\quad c_2 + c_1
\frac{(A - B)}{2} =0
\end{equation}
for the coefficients introduced in (\ref{7} - \ref{8}) are
satisfied, then deformed variations of the first two terms in
$\Gamma_{(4)}$ can cancel the last term in (\ref{3}). The
variation of the first complement $\delta_{0}^{({\cal W})} {\cal L
}_{(6|1)}$ (\ref{4}) in ${\cal W}$ under classical transformation
rules (\ref{1}) is
\begin{equation}\label{9} \delta_0^{({\cal W})} {\cal
L}_{(6|1)}=4\cdot\frac{5}{3}\frac{q^+q^-}{{\cal W}^2\bar{\cal
W}^4}q^{+a}(\bar{\varepsilon}^{\dot{\alpha}}_a
\bar{D}^-_{\dot{\alpha}}\bar{\cal W}+\varepsilon^{\alpha}_a
D^-_{\alpha}{\cal W}) D^4\ln{\cal W}
\end{equation}
$$
+\frac{4}{3}\cdot\frac{-2q^+q^-}{{\cal W}^2\bar{\cal W}^4}
q^{+a}\varepsilon^{\alpha}_a D^-_{\alpha}{\cal W}D^4\ln{\cal W}
+\frac{4}{3}\cdot\frac{\bar{\varepsilon}^{\dot\alpha}_a
\bar{D}^+_{\dot\alpha}q^{-a}}{{\cal W}^2\bar{\cal W}^3}(q^+ D^+
q^-)\frac{1}{16}D^+_{\alpha}D^{-2}\ln{\cal W}.
$$
Now introduce the second complement
\begin{equation}\label{14}
{\cal L}_{(6|2)}= d_2 [X^2\cdot\frac{1}{\bar{\cal W}^2}D^4 \ln
{\cal W} + X^2\cdot\frac{1}{{\cal W}^2}\bar{D}^4 \ln \bar{\cal
W}].
\end{equation}
Its variation in $q^{\pm}$ is exactly the first term in (\ref{9}),
and if we choose $d_2= -\frac{5}{3}$, it will cancel the
variations involving $\bar\varepsilon$. At the same time, the rest
of variation (\ref{9}) is saved. In order to cancel its first
part, one can consider variation of (\ref{5}) under the following
deformation of the transformations
\begin{equation}\label{12}
\delta_{(1|1)} \bar{\cal W} = \frac{{A}_{1}}{2}\cdot X
\varepsilon^{\alpha a}D^-_{\alpha} q^+_a\cdot \frac{1}{\bar{\cal
W}^2} D^4 \ln{\cal W},\ \delta_{(1|1)} {\cal W} =
\frac{\bar{A}_{1}}{2}\cdot X \bar\varepsilon^{\dot\alpha
a}\bar{D}^-_{\dot\alpha} q^+_a\cdot \frac{1}{{\cal W}^2} \bar{D}^4
\ln\bar{\cal W},\end{equation}
\begin{equation}\label{13} \delta_{(1|1)} q^-_a
= \frac{B_1}{4}\cdot X \varepsilon^{\alpha}_aD^-_{\alpha}{\cal W
}\frac{1}{\bar{\cal W}^2} D^4 \ln{\cal W}
  +\frac{\bar{B}_1}{4}\cdot X \bar\varepsilon^{\dot\alpha}_a\bar{D}^-_{\dot\alpha}\bar{\cal W
}\frac{1}{{\cal W}^2} \bar{D}^4 \ln\bar{\cal W}.
\end{equation}
Using these deformations, one can find variation of (\ref{5}): the
first term under $\delta^{(\bar{\cal W})}_{(1|1)}$, the second
term under $\delta^{(\bar{\cal W})}_{(1|0)}$ as well as
$\delta_{(1|1)}^{(q)}$, and the third term under
$\delta_{(1|0)}^{(q)}$. The part of the variations, we are
interested in, is
\begin{equation}\label{19}
\delta{\cal L}_{(4)} = c_1 \left[-\frac{2}{3}A_1 +B_1
+\frac{B-A}{3} \right](q^+q^-)q^{+a}\varepsilon^{\alpha}_a
D^-_{\alpha}{\cal W}\frac{D^4\ln{\cal W}}{{\cal W}^2\bar{\cal
W}^4}.
\end{equation}
The requirement of the cancellation between the second term in
$\Delta{\cal L}_{(6|1)}$ (\ref{9}) and the corresponding term
$\delta{\cal L }_{(4)}$ (\ref{19}) gives the following equation
\begin{equation}\label{coef0}
4\frac{2}{3}c_2=c_1(\frac{B-A}{3} +B_1-\frac{2}{3}{A}_1),\quad
{\rm or}\quad c_1(B_1-\frac{2}{3}A_1)=2c_2.
\end{equation}
These relations define coefficients in the decomposition of the
modified supersymmetry transformations in powers of $X$. There is
arbitrariness in choosing $A, B$, and, therefore, some additional
information is needed to get rid off this arbitrariness. For
cancelling the second part of (\ref{9}), we introduce a complement
of the new type ${\cal L}'_{(6|1)}= -\frac{1}{3}\cdot
X\cdot(\frac{q^+D^{+ \alpha}q^-}{{\cal W}\bar{\cal
W}})\frac{1}{\bar{\cal W}^2} D^+_{\alpha} D^{-2}\ln{\cal W}$.

The example presented above shows that, as a matter of principle,
a complement for ${\cal L}_{(6|0)}\propto D^4$ (\ref{2}) is
defined by the classical transformations (\ref{1}) generated by
$\bar\varepsilon$, while all contradictions arising in the
$\varepsilon$ sector variations can be eliminated by the hidden
transformation modifications $\delta_{(1|n)} \propto X^n$ order by
order. Thus, the the problem is split into two separate tasks.
This is our main idea how to overcome the difficulty in
constructing of the hidden ${\cal N}=2$ invariants with
derivatives of the vector strength of the ${\cal N}=2$ multiplet.

Let's consider a particular task of obtaining the leading term in
the full $F^6$ effective action, which can be solved with the
above considerations. For this purposes, it is sufficient to
consider a generic term of the series of complements to (\ref{2})
in the form
\begin{equation}\label{21}
\Gamma_{(6| n)}= d_n \int d^{12}z\,\left(\frac{-2 q^+ q^- }{{\cal
W}\bar{\cal W}}\right)^n \frac{1}{\bar{\cal W}^2} D^4 \ln {\cal W}
+ c.c.
\end{equation}
The classical variation $\delta_{0} {\cal L}_{(6|n)}$, generated
to the parameter $\bar\varepsilon$, for terms $\propto D^4\ln
{\cal W}$ is
\begin{equation}
\left[-d_n\cdot n\cdot\frac{(-2q^+q^-)^{n-1}}{{\cal W}^n\bar{\cal
W}^{n+2}} +
d_n\cdot\frac{n(n+4)}{n+2}\cdot\frac{(-2q^+q^-)^n}{{\cal
W}^{n+1}\bar{\cal W}^{n+3}}\right] \cdot
(q^{+a}\bar\epsilon^{\dot\alpha}_a \bar{D}^-_{\dot\alpha}\bar{\cal
W}) D^4\ln {\cal W}~.
\end{equation}
The requirement of cancellation variations of $ \delta_{0}{\cal
L}_{(6| n)}$ and $ \delta_{0}{\cal L}_{(6|n+1)}$ is fulfilled if
\begin{equation}
d_n = d \frac{(n+2)(n+3)}{n}.
\end{equation}
Summing all complements ${\cal L}^q_{(6)}=
\sum_{n=0}^{\infty}{\cal L}_{(6| n)}(X)$, one obtains
\begin{equation} \Gamma^q_{(6)}=-\frac{c_2}{6}\int
d^{12}z\,[\frac{X}{(1 - X)^2} + \frac{5 X}{1 - X} - 6\ln (1 - X)
]\frac{1}{\bar{\cal W}^2} D^4 \ln {\cal W}.
\end{equation}
Of course, this result is not entirely full in the sense that it
should be completed by contributions containing hypermultiplet
derivatives. Generic terms with $q^+D^+q^{-}$ derivatives can be
found.  We introduce a new type of complement
\begin{equation}
\label{57}{\cal L}'_{(6|n)} = p_n \frac{(-2q^+q^-)^n}{{\cal
W}^{n+1}\bar{\cal W}^{n+3}}q^+D^+_{\alpha}q^- D^{+\alpha}
D^{-3}\ln{\cal W}.
\end{equation}
Then the requirement of the cancellation of its variation
$\delta^{q}_0(\bar{\varepsilon})$ with an appropriate term in
variation (\ref{21}) leads to a inhomogeneous recurrent relation.
The relation has the solution
\begin{equation}\label{solv}
p_n =-\frac{1}{6}\cdot\frac{(n+2)(n+3)}{n+1}
-\frac{1}{12}\cdot\frac{(n+2)^2(n+3)}{n+7}H^{(2)}_{n+1}~,
\end{equation}
where $H^{(2)}_n =\sum_{k=1}^{n}{1\over k^{6}}$ is the harmonic
number $[n,2]$. For the coefficients of the next generic term of
the series ${\cal L}''_{(6|n)} =
h_{n}X^{n}(q^{+}D^{+}q^{-})^{2}D^{-2}\ln {\cal W}$, there must be
the recurrent relation on $p_{n}, d_{n}, h_{n}$.

We see that for ${\cal N}=4$ supersymmetrization of
next-to-leading terms, one has to consider transformations that
mix different terms in the derivative expansion of the effective
action! To make sure that the guessed transformations
(\ref{7}-\ref{8}) are not unreasonable, one can consider a
variation of the well-known classical action \cite{12}. The
variation of the hypermultiplet action is proportional to the
on-shell equation of motion $D^{++}q^{+}=0$, but variation of the
vector strength action in (\ref{7}) is
\begin{equation}\label{deltcl} \delta_{(1|0)}
\Gamma_{0}=\frac{1}{8}\int d^{12}z
\{\bar{A}\bar{\varepsilon}^{\dot\alpha
a}\bar{D}^-_{\dot\alpha}q^+_a \frac{1}{{\cal W}}\ln\bar{\cal W} +
A\varepsilon^{\alpha a}D^-_{\alpha} q^+_a \frac{1}{\bar{\cal
W}}\ln {\cal W} \}~.
\end{equation}
This expression has the same structure as $\delta_0 {\cal
L}_{(4|0)} $. The fact that in order to cancel (\ref{deltcl}) one
should take into account classical variations of ${\cal
L}_{(4|0)}\propto \ln{\cal W}\ln\bar{\cal W}$ serves as an
additional cross-checking of the consistency of the proposed
recipe.

It is also interesting to consider the first deformed variations
of $F^6$ (\ref{2}). We obtain
\begin{equation}\label{del6}
\delta_{(1|0)}{\cal L}_{(6|0)} = c_2 [\frac{\bar{A}}{{\cal
W}^3\bar{\cal W}^2} {\bar{\varepsilon}}^{\dot{\alpha} a}
\bar{D}^-_{\dot{\alpha}} q^+_a D^4 \ln {\cal W} \bar{D}^4\ln
\bar{\cal W} - \frac{{A}}{\bar{\cal W}^5} \varepsilon^{\alpha a}
D^-_{\alpha} q^+_a \ln{\cal W} (D^4 \ln {\cal W})^2]~.
\end{equation}
The first term  in the brackets looks like classical variation of
the one-loop $F^8$ structure ${\cal
L}_{(8|0)}={\bf\Psi}^2\bar{\bf\Psi}^2=\frac{1}{{\cal
W}^2}\bar{D}^2\ln\bar{\cal W}\frac{1}{\bar{\cal W}^2}D^4\ln{\cal W
}$ but with coefficient $c_2$. This means that the one-loop
coefficient ${1\over 2(24\pi)^2}$ in front of $F^8$ structure
\cite{4},\cite{21} of the effective action will be renormalized by
two-loop contributions. The second term in (\ref{del6}) is a term
of new type. For its cancellation, one needs to add another
structure
\begin{equation}
\Gamma_{(8|0)}'= c_8 \int d^{12}z\,\ln {\cal W}
\left(\frac{1}{\bar{\cal W}^2} D^4\ln {\cal W}\right)^2~,
\end{equation}
with typical $\propto \frac{1}{(4\pi)^4}$ two-loop coefficient
$c_8 = -c_2\frac{A}{2}$. This allows us to conjecture that
accurate two-loop calculations should give such $F^{8}$-structure
in the effective action.

Thus, the self-consistent obtaining of the appropriate
hypermultiplet dependent contributions in the effective action and
modification of the hidden supersymmetry transformation allows one
to obtain information about renormalizable terms as well as
non-renormalizable higher corrections terms.

\section{Acknowledgements}
The authors thank the Organizing Committee of the SQS'03
conference for warm welcome and partial support. The work was
supported in part by INTAS grant, INTAS-00-00254 and RFBR grant,
project No 03-02-16193. The authors also would like to thank
I.L.Buchbinder and E.A. Ivanov for numerous discussion and a
critical remarks.

\end{document}